# Three-Dimensional Yielding in Anisotropic Materials: Validation of Hill Criterion

Manish Kaushal[1] and Yogesh M Joshi[2]

[1]Department of Chemical Engineering, Indian Institute of Technology Kharagpur

[2]Department of Chemical Engineering, Indian Institute of Technology Kanpur

Corresponding authors:

MK, E Mail: mkaushal@che.iitkgp.ac.in; YMJ, E Mail: joshi@iitk.ac.in

## Abstract

Yielding transition in isotropic soft materials under superposition of orthogonal deformation fields is known to follow von Mises' criterion. However, in anisotropic soft materials von Mises' criterion fails owing to preferred directions associated with the system. In this work we study a model anisotropic yield stress system: electrorheological (ER) fluids that show structure formation in the direction of electric field. We subject the ER fluids to superposition of orthogonal stress fields that leads to different yield stress values. We obtain a yielding state diagram by plotting normalized rotational shear stress against normalized radial shear stress corresponding to yield point for a given electric field. Remarkably, the state diagram validates the Hill's yielding criterion, which is a general yielding criterion for materials having anisotropy along three orthogonal directions, originally developed for metallic systems. Validation of Hill's criterion suggests the universality of its application to anisotropic systems including conventional anisotropic soft materials having yield stress.

## I. Introduction:

Soft materials such as colloidal suspension, emulsions, physical or chemical gel, cosmetic/pharmaceutical pastes and creams, bio-fluids, slurry, granular materials, food products, etc. are ubiquitous in nature and industry. One important class of such materials are *structurally arrested soft materials*, wherein constituent elements possess very low translational mobility over the mesoscopic length-scales.[1-3] The origin of such structural arrest could be attractive (depletion, electrostatic, hydrophobic/philic, van der Waals interactions, electric/magnetic field-induced dipolar-multipolar interactions, etc.) or repulsive (excluded volume, electrostatic, etc.)[4]. Characteristically these materials tend to undergo predominantly an elastic deformation at finite magnitudes of stress. On the other hand, above a threshold value of stress that is sufficient to break the structure, material shows dissipative flow. Over a century ago this aspect was articulated by Bingham,[5] wherein he recognized the necessity of overcoming friction in 'structured materials', before they undergo a plastic flow. Such threshold stress required to facilitate plastic flow in soft solid-like materials has been termed as yield stress. In general, the understanding of yielding phenomenon that triggers solid – liquid transition under the stress field is of great academic and industrial importance. Moreover, since in most of the practical applications these materials experience simultaneous imposition of multiple deformation fields, the yielding is needed to be understood from a three-dimensional perspective. For isotropic materials, when material is subjected to superposition of multiple deformation fields, plastic flow has been observed to occur when second invariant of the stress tensor exceeds the value of yield stress, thereby validating the von Mises' criterion.[6] However, for the anisotropic systems, the yield stress is expected to be dependent upon the direction of shear vis a vis preferred direction of the structure in a material. Under such conditions the von Mises' criterion is expected to fail. In this work we investigate yielding in electrorheological fluids that show electric field-induced anisotropic microstructure and therefore direction dependent yield stress. We observe that such system validates Hill criterion.[7-8]

For isotropic materials that show yield stress, the constitutive equation relating deviatoric stress tensor ($\underset{\sim}{\tau}$), strain tensor ($\underset{\sim}{\gamma}$) and rate of strain tensor ($\underset{\sim}{\dot{\gamma}}$) is given by:[9-10]

$$\underset{\sim}{\tau} = G\underset{\sim}{\gamma}, \qquad \text{for } \bar{\tau} < \tau_y$$

$$\underset{\sim}{\tau} = \left(\frac{\tau_y}{\dot{\gamma}} + \eta\right)\underset{\sim}{\dot{\gamma}}, \qquad \text{for } \bar{\tau} \geq \tau_y, \qquad (1)$$

where $G$ is the elastic modulus, $\eta$ is the deformation field dependent viscosity and $\dot{\gamma}$ is the second invariant of the rate of strain tensor. The von Mises' yielding criterion suggests that material undergoes plastic flow only when magnitude of stress ($\bar{\tau}$), given by second invariant of the stress tensor:

$$\bar{\tau} = \sqrt{\underset{\sim}{\tau} : \underset{\sim}{\tau}/2}, \qquad (2)$$

exceeds the yield stress ($\tau_y$). For a constant value of $\eta$, equation (1) denotes a Bingham model, while for $\log(\eta) \propto \log(\dot{\gamma})$, we get the Herschel-Bulkley model.[11] Whether viscous dissipation is strictly prohibited for $\bar{\tau} < \tau_y$ or not is a perennial question and out of the scope for the present work. Although there is a large body of the literature that supports the existence of engineering and/or true yield stress, a threshold stress above which material undergoes noticeable plastic deformation, the experimental validation of von Mises' criterion was carried out for soft materials rather recently by Ovarlez and coworkers.[12] They studied three - dimensional yielding for variety of structurally arrested soft materials under simultaneous imposition of combination of rotational and squeeze flow. Following the same lines, Shaukat and co-workers studied yielding of variety of soft materials under superposition of rotational and radial shear flow field (induced by tensile force).[10] Both the groups successfully validated the von Mises' yielding criterion for isotropic materials.

von Mises' criterion, however, does not work for anisotropic materials that have different yield stresses in the different planes of deformation. To address this problem Hill proposed a general yield criterion for materials having anisotropy along three mutually orthogonal directions (known as orthotropic materials), and is given by equation:[7-8]

$$F(\tau_{rr} - \tau_{zz})^2 + G(\tau_{zz} - \tau_{\theta\theta})^2 + H(\tau_{\theta\theta} - \tau_{rr})^2 + \frac{\tau_{rz}^2}{\tau_{rz,y}^2} + \frac{\tau_{\theta z}^2}{\tau_{\theta z,y}^2} + \frac{\tau_{r\theta}^2}{\tau_{r\theta,y}^2} = 1, \quad (3)$$

where $\tau_{ij}$ are the different components of the stress tensor in the polar coordinate system. The parameters $F$, $G$, and $H$ are dependent on the yield stresses in different directions given by: $2F = \tau_{rr,y}^{-2} + \tau_{zz,y}^{-2} - \tau_{\theta\theta,y}^{-2}$, $2G = \tau_{zz,y}^{-2} + \tau_{\theta\theta,y}^{-2} - \tau_{rr,y}^{-2}$, and $2H = \tau_{\theta\theta,y}^{-2} + \tau_{rr,y}^{-2} - \tau_{zz,y}^{-2}$.[7] The Hill yield criterion given by Eq. (3) reduces to von Mises' criterion given by Eq. (2) in a limit of vanishingly small anisotropy. Hill originally defined a quadratic yield criterion for orthotropic metal sheets having in-plane isotropy (also known as normal anisotropy),[8] he further extended this criterion to materials containing in-plane anisotropy, so that it can be applied to any kind of orthotropic material.[13] Later on various modifications have been made to Hill's criterion, and have been extensively used to study the yielding phenomenon for a broad class of metal/alloy sheets. Badr et al. implemented Hill's extended criterion for Ti alloy sheets having in-plane anisotropy,[14] Ghennai et al. used this criterion for identification of plastic anisotropy in steel metal sheets,[15] Cardoso et. al. modified Hill's criterion by incorporating biaxial symmetric stress tensor and implemented it to study the yielding behavior of Aluminum alloy sheets.[16]

While validations of Hill criterion for orthotropic metal/alloy sheets are available in literature, yielding behavior of anisotropic soft materials has not yet been investigated under different deformation fields. One such interesting system is electrorheological (ER) fluids, which are dispersions of colloidal dielectric particles in a non-conducting solvent. The structural arrest in these systems occurs due to electric field-induced dipolar/multipolar interaction.[17-19] Since the particle-dipoles are aligned in the direction of applied electric field, the strength of material is highest in the direction of field as compared to any other direction, which makes these jammed suspensions anisotropic. When such an electrically jammed ER suspension is subjected to deformation field, in the limit of very small stress/strain field, it responds elastically. However, once the applied deformation field exceeds the yield stress, particle-chains start breaking, which leads to structural breakdown.[18-21] Such electric field-controlled fluidity makes these fluids a potential candidate for many practical applications such as, breaks, clutches, dampers etc.[22-23]

Many experimental as well as simulation studies have been performed to understand the yielding phenomenon in ER fluids, however most of such studies were performed in unidirectional flow modes such as, purely rotational, purely squeeze, or purely tensile/elongation flow mode. Tian et al.[24] performed experiments on ER suspension in rotational shear, elongation, and compressional flow modes independently and measured the yield stress. They found that yield stress during compression is higher than both tensile and shear yield stresses, whereas the tensile yield stress is three to four times greater than the shear yield stress. They also studied the confinement effect and found that yield stress values for relatively smaller initial gaps (<2mm) differ significantly as compared to large gaps, which is attributed to strength dependence on chain-length and its cross-sectional diameter.[25] Meng and coworkers[26] measured flow resistance of ER fluids under unidirectional compression along the electric field direction and observed that compressive resistance values are significantly higher as compared to the theoretical estimate of non-Newtonian squeeze flow theory. In this work, we study the yielding of two types of ER suspensions under simultaneous application of rotational shear flow field and radial shear flow field, wherein the former one is induced by applying torque on the upper plate and the later one by application of upward normal force on the same. The ER suspension under electric field shows many similarities with conventional yield stress materials. We therefore believe that the results presented in this work have broader implications for anisotropic yield stress materials in general.

**II. Materials and Experimental Methods:**

In this work we use two ER fluids: 34 mass % Corn-starch (LobaChemie)-silicon oil and 17 mass % Zeolite (Sigma Aldrich)-silicon oil suspension. The powder form of both materials was dispersed in silicon oil and stirred for 30 minutes using mechanical stirrer to prepare the stable ER suspensions. All the experiments were conducted using Anton-Paar MCR501 rheometer attached with Electrorheology set-up. The suspension was kept between the parallel plates (diameter: 50mm) with an initial gap of 1 mm, and the electric field was applied across the plates in the negative $z$-direction, as shown in figure 1. At first the sample was placed without any electric field, for which the normal force recorded by the rheometer is practically zero. As soon as the electric field was switched on, rheometer records very small normal force induced by the material on the top plate due to electric field-induced structure formation, whose contribution towards yielding is negligible. We kept the loading uniform throughout our experiments, and followed the identical equilibration procedure for every experiment. A torque ($T$) was applied on the upper plate to induce the rotational shear flow, whereas an upward normal ($F_N$) was exerted on the same to induce the radial shear flow, as shown in figure 1.

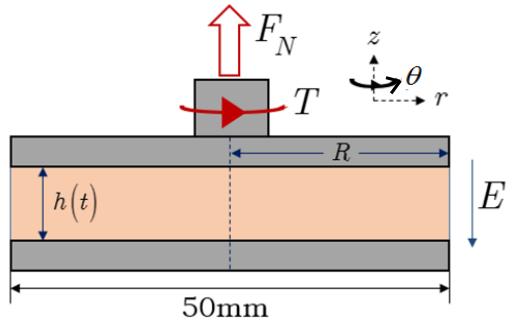

**Figure 1:** Schematic of a parallel plate geometry employed in the present work.

In combined flow mode both $T$ and $F_N$ are imposed simultaneously. We use polar co-ordinate system, wherein $r$, $\theta$, and $z$ represent radial, azimuthal/angular and axial directions respectively. In tensile mode and combined mode experiments, the gap between plates always remains significantly smaller as compared to plate radius ($h(t) << R$) up to the yield point, which enables us to use lubrication approximation.[27] Therefore, we have two in-plane shear stresses acting on $r - \theta$ surface: rotational shear stress ($\tau_{\theta z}$) in $\theta$ direction induced by rotational torque, and radial shear stress ($\tau_{rz}$) induced by normal force that causes flow in the radially inward direction.

**III. Results and discussion:**

As mentioned in the previous section, we perform the experiments in three modes: tensile mode by applying normal force ($F_N$) on the upper plate, rotational shear mode by applying torque ($T$) on the upper plate, and combined mode wherein superposition of upward normal force and torque is applied on the top plate.

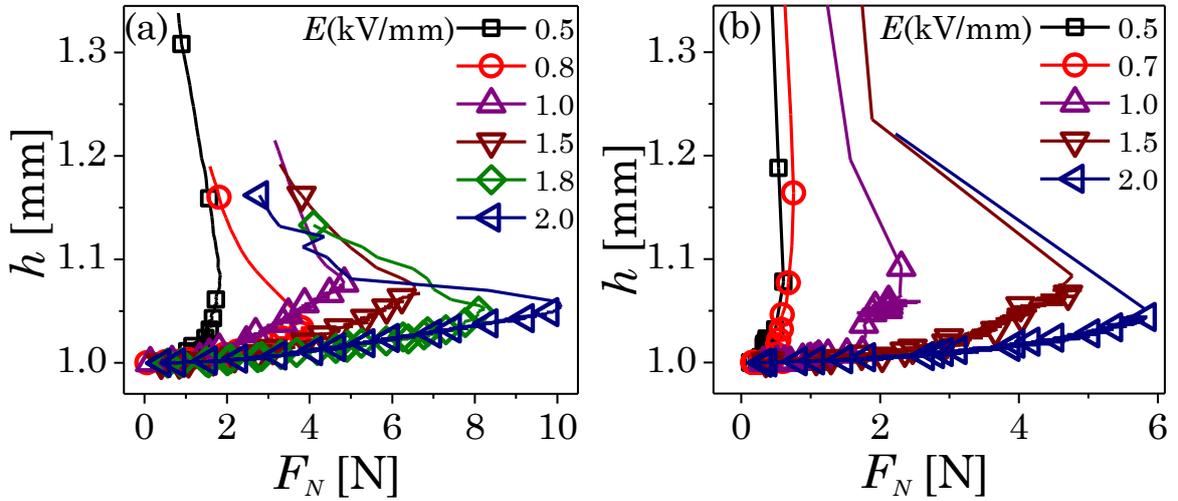

**Figure 2:** Gap ($h$) between the two parallel plates as a function of upward normal force ($F_N$) imposed on the upper plate in purely tensile mode at various electric field strengths ($E$) for Corn-starch suspension in (a), and for Zeolite suspension in (b). While applying normal force ramp, $F_N$ reaches up to a certain maximum value, after which material fails, resulting an abrupt decrease in $F_N$ (as rheometer is no longer able to control the ramp), this point is considered as the yield point.

In purely tensile mode a linear ramp of upward normal force ($F_N$) is imposed on the top plate for a given electric field ($E$). Figures 2(a) and (b) show gap ($h$) as a function of $F_N$ for Corn-starch suspension and Zeolite suspension respectively. It can be seen that with increase in $F_N$, $h$ changes linearly, which manifests the elastic regime. It is followed by a slight deviation from linearity, wherein $h$ increases faster than linear with increase in $F_N$. Eventually, the material fails suddenly that causes an abrupt decrease in $F_N$ as rheometer is no longer able to control the ramp. Accordingly, the point corresponding to maximum value of $F_N$ achieved, shown in figure 2, is considered as the yield point for a given $E$. Such sudden decrease in normal force is due to structural breakdown of the whole bulk material taking place at the yield point. It can be seen that with increase in $E$, the extent of polarization caused by electric field increases, leading to augmentation of the attractive forces of larger magnitude among the particles, consequently $h$ decreases with $E$ for a constant value of $F_N$.

It is apparent from figure 2 that up to the yield point, change in gap (5-7% of the initial gap) is very small. Consequently, the electric field practically remains constant till the material failure point. Accordingly, the ratio of the inter-plate gap ($h(t)$) to the radius of plate ($R$) also remains significantly small ($h(t) << R$) up to the yield point, that results in $v_z << v_r$, $\frac{\partial v_z}{\partial z} << \frac{\partial v_r}{\partial z}$, where $v_z$ and $v_r$ are the axial and radial velocity components respectively. In addition, any other velocity gradients in $\theta$ and $r$ direction are also negligible. Owing to the above approximation (known as lubrication approximation) application of normal force leads to negligible normal stress but finite radial shear stress ($\tau_{rz}$).[28-29] Therefore, in the tensile mode experiments, material undergoes predominantly the radial shear flow.[10,12,30] The corresponding (inward) radial shear stress ($\tau_{rz}$) can be calculated by using lubrication approximation theory,[27] and is given by:

$$\tau_{rz} = \frac{3F_N h}{2\pi R^3}. \qquad (4)$$

The radial shear yield stress ($\tau_{rz,y}$), therefore, corresponds the maximum normal force shown in figure 2. In figure 3, $\tau_{rz,y}$ is plotted as a function of $E$ for the Cornstarch and Zeolite suspensions. As expected, we observe a monotonic increase in $\tau_{rz,y}$ with $E$ due to enhancement in polarization forces. For both the suspensions, $\tau_{rz,y}$ demonstrates a power law dependence on $E$ with the cornstarch suspension ($\tau_{rz,y} \sim E^{1.1}$) showing weaker dependence than that of for the Zeolite suspension ($\tau_{rz,y} \sim E^{1.8}$).

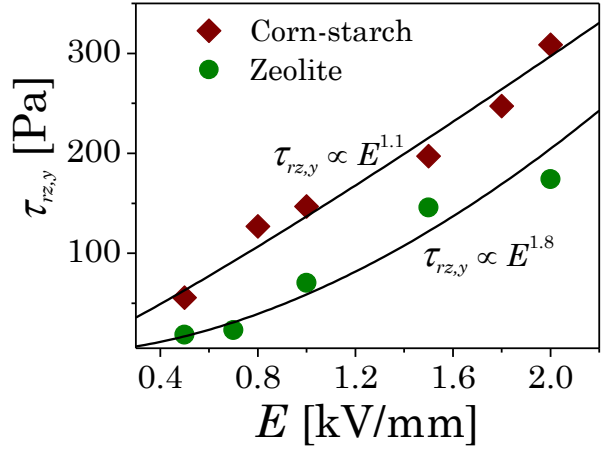

**Figure 3:** Radial shear yield stress ($\tau_{rz,y}$) is plotted as a function of $E$. The solid lines show power law fits (Corn-starch suspension: $\tau_{rz,y} \propto E^{1.1}$, and Zeolite suspension: $\tau_{rz,y} \propto E^{1.8}$). $\tau_{rz,y}$ is calculated using equation (4), wherein maximum value of $F_N$ achieved in tensile mode experiment for a given $E$ (as shown in figure 2) is used.

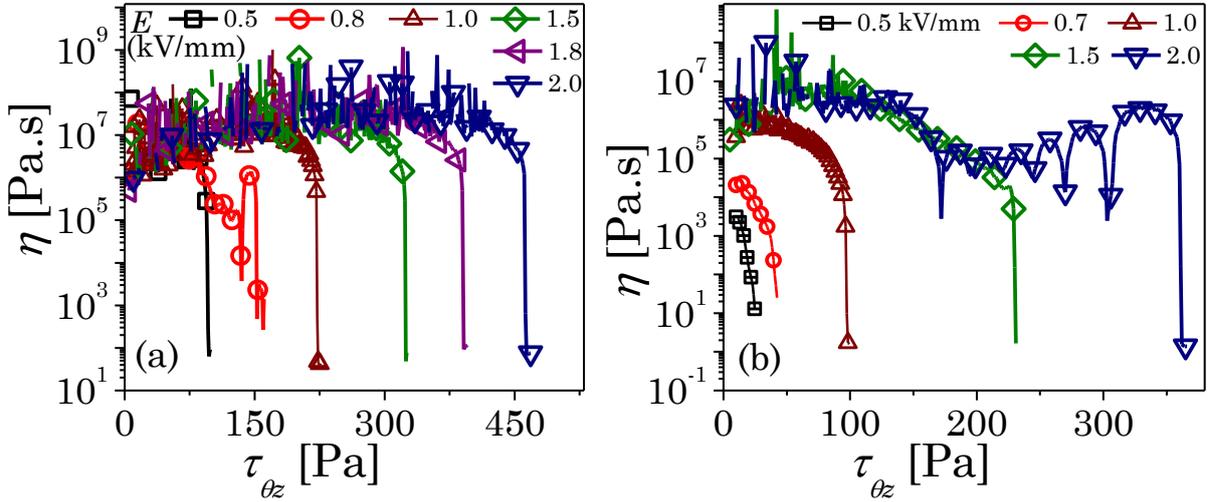

**Figure 4:** Viscosity ($\eta$) is plotted as a function of rotational shear stress ($\tau_{\theta z}$) in the purely rotational shear flow mode at different $E$ for Corn-starch suspension in (a), and for Zeolite suspension in (b). The rotational shear stress associated with sudden drop in viscosity corresponds to rotational shear yield stress ($\tau_{\theta z,y}$) at a given $E$.

In purely rotational flow mode, a torque is applied on the upper plate, which induces rotational shear flow in the material, the corresponding shear stress ($\tau_{\theta z}$) is directly given by the rheometer. Figure 4 shows results for rotational shear flow mode at different $E$ for the Corn-starch suspension in (a), and for the Zeolite suspension in (b), wherein a rotational shear stress ramp is imposed on the sample and viscosity is measured. As expected, viscosity drops abruptly at the material failure point, the shear stress associated with this point corresponds to the rotational shear yield stress ($\tau_{\theta z,y}$) for given $E$. Figure 5 shows $\tau_{\theta z,y}$ as a function of $E$ for the Corn-starch and Zeolite suspensions. For both the suspensions, $\tau_{\theta z,y}$ shows a power law dependence on $E$ (shown by solid lines, Corn-starch suspension: $\tau_{\theta z,y} \propto E^{1.1}$, Zeolite suspension: $\tau_{\theta z,y} \propto E^{1.9}$). Similar to that observed for $\tau_{rz,y}$, increase in $\tau_{\theta z,y}$ with $E$ is also due to greater extent of polarization forces at higher $E$, which leads to increase in inter-particle attractive forces.

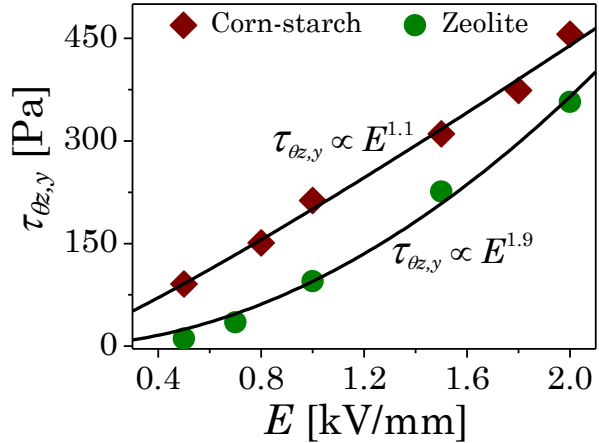

**Figure 5:** Rotational shear yield stress ($\tau_{\theta z,y}$) is plotted as a function of electric field strength ($E$). Solid lines show power law fits for Corn-starch suspension: $\tau_{\theta z,y} \propto E^{1.1}$, and Zeolite suspension: $\tau_{\theta z,y} \propto E^{1.9}$.

It is important to note that rotational shear yield stress ($\tau_{\theta z,y}$) is always higher than radial shear yield stress ($\tau_{rz,y}$) at any $E$. This behavior expectedly invalidates von Mises' yielding criterion proposed for isotropic materials. Furthermore, the various exponents observed for power law dependence of $\tau_{\theta z,y}$ and $\tau_{rz,y}$ on $E$ depend on the mechanism responsible for ER effect, which ultimately is related to the material properties. It is important to note that for a given ER fluid, the power law dependence of $\tau_{\theta z,y}$ and $\tau_{rz,y}$ on $E$ are almost identical, In the literature three models (mechanisms) have been proposed for the ER effect: the polarization model,[18,22] the conduction model,[31] and the saturation-surface-polarization model.[32] These models respectively lead to yield stress dependence on $E$ as: $\tau_y \propto E^2$, $\tau_y \propto E^{1.5}$ and $\tau_y \propto E$.[18,22,31] In the present work, Corn-starch suspension seems to follow the saturation-surface-polarization model, whereas Zeolite suspension suggests agreement with the polarization model. A slight deviation observed in the experiments from that of the theoretical prediction may be due to polydispersity and current leakage at the electrode plates.

Normal stress associated with the material can, in principle, have a profound effect on the yield stress behavior.[33] Therefore, it is important to discuss the effect of normal stress/force generated due to application of electric field and how does effect of the same compare with the stress at yield point. We discuss the case when there is no external application of normal force and the yielding is purely due to the rotational flow field. Supplementary information figures SI-1(a) and SI-2(a) show the normal force ($F_N$) experienced by the top plate of the rheometer as a function of rotational shear stress ($\tau_{\theta z}$) for Corn-starch suspension, at the electric field strengths, $E$ =0.5 and 1.0 kV/mm respectively. The corresponding shear rate has also been plotted on the right ordinate. It should be noted that, since no external normal force is applied on the top plate, the normal force ($F_N$) measured by the rheometer is due to the electric field induced structure formation in a limit $\tau_{\theta z} \to 0$ (no stretching of particle-chains). The additional normal force at higher $\tau_{\theta z}$, on the other hand, is due to stretching of the electric field induced particle chains before the material yields. In order to evaluate the actual contribution of the above mentioned induced normal force towards yielding, we also plot the radial shear stress ($\tau_{rz}$, calculated using $F_N$) as a function of rotational shear stress ($\tau_{\theta z}$) near the yield point during the purely rotational experiments for the Corn-starch suspension, in the supplementary information figures SI-1(b) and SI-2(b) for $E = 0.5$ and 1.0 kV/mm respectively. It can be seen that near the yield point, $\tau_{rz}$ values are significantly smaller than the imposed rotational shear stress values ($\tau_{\theta z}$). In the inset of the supplementary information figures SI-1(b) and SI-2(b), we also plot $\tau_{rz}/\tau_{rz,y}$ versus $\tau_{\theta z}/\tau_{\theta z,y}$, which convincingly shows normal force $F_N$ (or $\tau_{rz}$) has negligible contribution towards yielding. We also observe similar behavior for the other explored electric field strengths in both the studied systems.

We have also checked the contribution of shear thickening on the observed behavior. We plot the rotational shear rate ($\dot{\gamma}$) up to the yield point as a function of imposed rotational shear stress ($\tau_{\theta z}$) for purely rotational flow case, as shown in supplementary information figures SI-1(a) and SI-2(a). Since this is a pre-yielding regime, the induced shear rate is extremely small (of the order $10^{-7}$-$10^{-5}$) for shear thickening to occur. (Please note that Fall et al.,[34] who studied 41 weight % Corn-starch suspension, reported shear rate of the order of $10^0$ at the onset of shear thickening). Furthermore, for the electro-rheological fluids under electric field the constituent dielectric particles are trapped due to electrostatic forces and hence there is no chance of any sedimentation, even less below the yield stress.

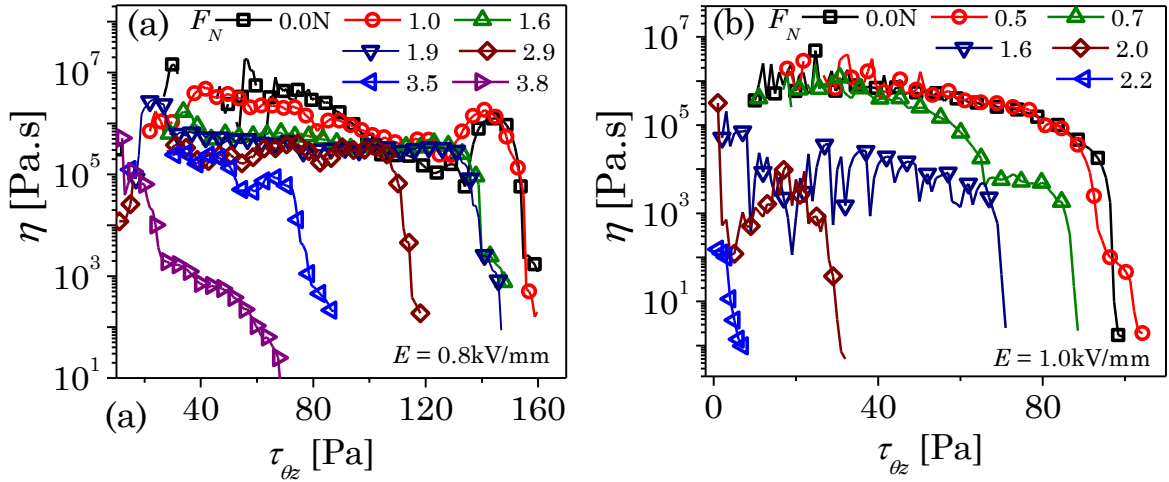

**Figure 6:** Viscosity ($\eta$) is plotted as a function of rotational shear stress ($\tau_{\theta z}$) in combined flow mode experiment (under simultaneous application of $\tau_{\theta z}$ ramp at a constant $F_N$ imposed on top plate in upward direction) for Corn-starch suspension at $E = 0.8$ kV/mm in (a), and for Zeolite suspension at $E = 1.0$ kV/mm in (b). Such experiments have been conducted at other electric fields also for both suspensions.

We now investigate the yielding behavior under superposition of orthogonal deformation fields, wherein an ER suspension under constant electric field is subjected to rotational shear stress ($\tau_{\theta z}$) ramp under application of constant normal force ($F_N$). In these experiments we make sure that $\tau_{rz}$ value corresponding to applied $F_N$ should not be greater than $\tau_{rz,y}$ for a given $E$ (so that yielding does not take place only due to $F_N$, without any contribution from $\tau_{\theta z}$). Figure 6 shows shear viscosity ($\eta$) as a of function of $\tau_{\theta z}$ for different values of $F_N$ for Corn-starch suspension at $E = 0.8$ kV/mm in (a), and for Zeolite suspension at $E = 1.0$ kV/mm in (b). The point, at which viscosity drops abruptly, corresponds to the yield point. Accordingly, for each such yield point we obtain $\tau_{rz}$ and $\tau_{\theta z}$. It can be seen that with increase in magnitude of $F_N$, the yield point shifts towards lower $\tau_{\theta z}$ value. This suggests that as the contribution of normal force towards unjamming increases, lesser contribution of rotational shear stress is required to yield the material.[35-36] We perform the above set of experiments for constant values of $E$ over a range: 0.5 – 2 kV/mm for both the suspensions and obtain the yield stress data.

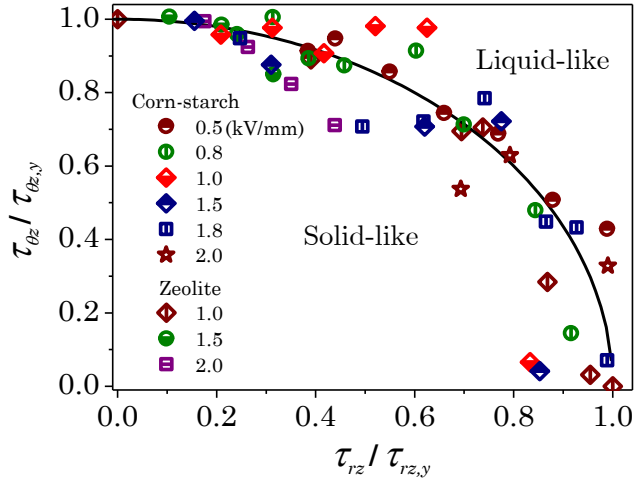

**Figure 7:** A yielding state diagram is created using rotational and radial shear stress values corresponding to yield point in combined flow experiment. Normalized rotational shear stress ($\tau_{\theta z}/\tau_{\theta z,y}$) is plotted against normalized radial shear stress

($\tau_{rz}/\tau_{rz,y}$) for the Zeolite and Corn-starch suspensions at different electric fields ($E$) ranging from 0.5 to 2.0kV/mm The solid line represents the curve generated by equation (5), which is a two dimensional form of *Hill's yielding criterion.*

We are now in position to assess the validity of Hill's criterion. Since for the present case only two in-plane orthogonal shear stresses $\tau_{\theta z}$ and $\tau_{rz}$ act on the material, the Hill's criterion reduces to:

$$\left(\frac{\tau_{rz}}{\tau_{rz,y}}\right)^2 + \left(\frac{\tau_{\theta z}}{\tau_{\theta z,y}}\right)^2 = 1 \qquad (5)$$

Accordingly, in figure 7, we plot the normalized rotational shear stress ($\tau_{\theta z}/\tau_{\theta z,y}$) against the normalized radial shear stress ($\tau_{rz}/\tau_{rz,y}$) corresponding to the point of yield. It can be seen that the experimental data, irrespective of the magnitude of electric field and the nature of a system, falls within a narrow range, suggesting a yielding boundary between solid-like (elastic) response to liquid-like (viscous) response. The solid line in figure 7 shows yielding surface/boundary resulted from eq. (5). Remarkably the experimental yield stress data in the combined flow mode lie in very close proximity to the yielding boundary validating Hill's yielding criterion for ER fluids under electric field. It can be seen that the yielding state diagram shown in figure 7 is universally applicable to ER suspensions irrespective of their nature and magnitude of electric field. The region below the yielding phase line corresponds to the jammed/solid regime, whereas the region above the line represents fluidized/yielded state. We believe that such validation of universal yield criterion for ER fluids will provide a profound understanding of the yielding behavior of these fluids, which will be very useful while designing ER fluid-based devices on one hand and understanding of the yielding phenomenon of anisotropic yield stress soft materials in general on the other.

## IV. Conclusion:

In this work we study yielding behavior of an electrically jammed electrorheological fluid that forms a model anisotropic yield stress soft system, under superposition of rotational shear and tensile force-induced radial shear flow fields. To begin with we independently measure rotational and radial shear yield stresses for two types of ER suspensions. It is observed that the absolute values of the yield stresses in different directions vary significantly, thereby violating von Mises' yielding criterion meant for isotropic materials. The behavior is, therefore, attributed to the electric field-induced anisotropy. In case of combined flow experiments, we subject the system to superposition of normal force and rotational shear stress (torque). It is observed that with increase in magnitude of normal force, ER system yields at lower rotation shear stress; suggesting yielding is indeed a three-dimensional phenomenon. We observe that the Hill's yielding criterion proposed for the anisotropic materials shows remarkable agreement with the experimental data. We believe that the present work not just renders insight into the yielding behavior in the ER systems where cause of anisotropy is application of electric field but is also applicable to anisotropic yield stress soft materials in general.

**Acknowledgement:** The financial support from the Department of Atomic Energy–Science Research Council (DAE-SRC), Government of India is greatly acknowledged.

# Supplementary information

# Three-Dimensional Yielding in Anisotropic Materials: Validation of Hill Criterion

Manish Kaushal* and Yogesh M Joshi[1]

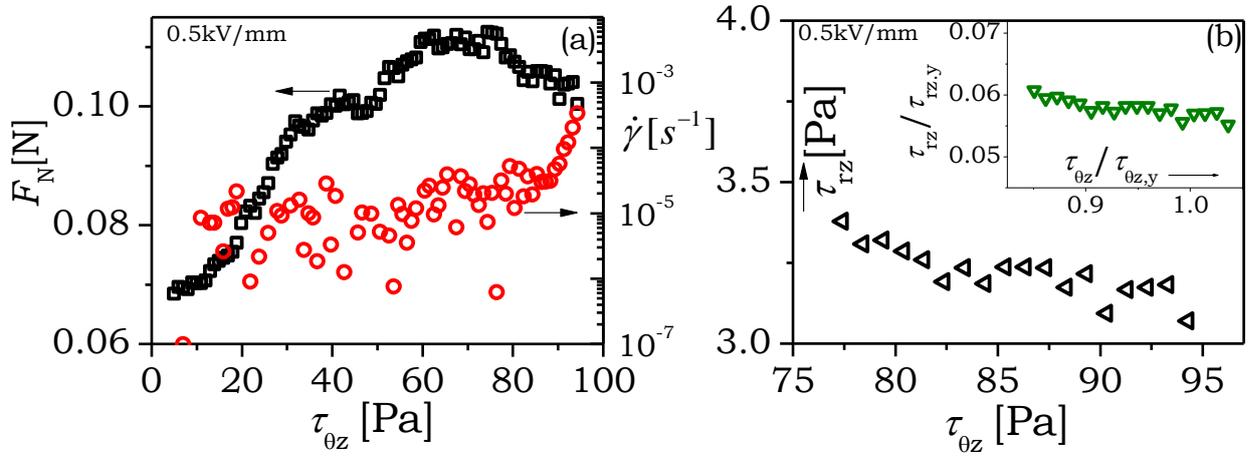

Figure SI-1: For the Corn-starch suspension, (a) Normal force ($F_N$) experienced by the rheometer top plate and the corresponding rotational shear rate ($\dot{\gamma}$) is plotted as a function of rotational shear stress ($\tau_{\theta z}$) till the yield point in the purely rotational experiment. For the same system, (b) Radial shear stress ($\tau_{rz}$) is plotted as a function of rotational shear stress ($\tau_{\theta z}$). In the inset of (b) $\tau_{rz}/\tau_{rz,y}$ is plotted with respect to $\tau_{\theta z}/\tau_{\theta z,y}$. All the above results are obtained at $E = 0.5$ kV/mm.

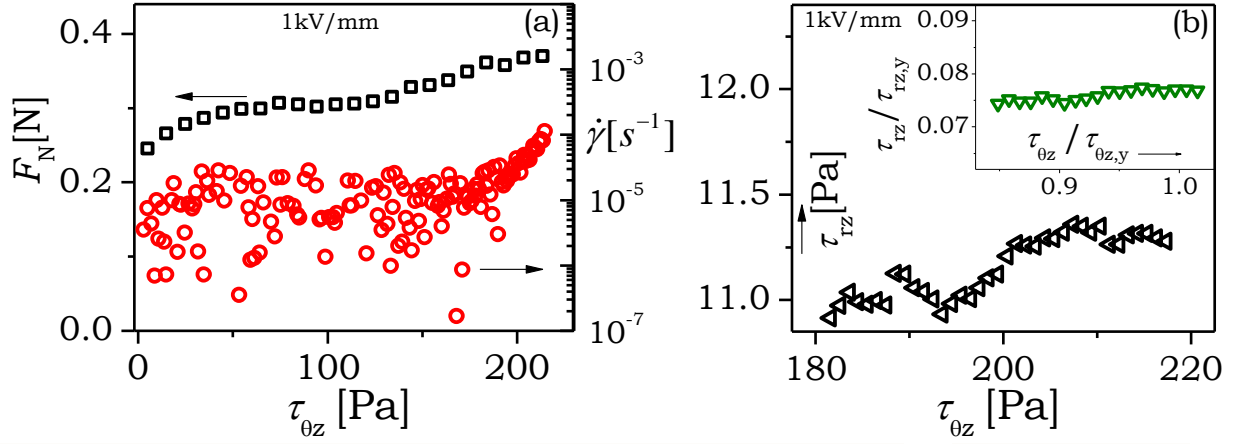

Figure SI-2: For the Corn-starch suspension (a) Normal force ($F_N$) experienced by the rheometer top plate and the corresponding rotational shear rate ($\dot{\gamma}$) is plotted as a function of rotational shear stress ($\tau_{\theta z}$) till the yield point in the purely rotational experiment. For the same system (b) Radial shear stress ($\tau_{rz}$) is plotted as a function of rotational shear stress ($\tau_{\theta z}$). In the inset of (b) $\tau_{rz}/\tau_{rz,y}$ is plotted with respect to $\tau_{\theta z}/\tau_{\theta z,y}$. All the above results are obtained at $E = 1.0$ kV/mm.